\documentclass{article}

\usepackage[final,nonatbib]{neurips_2020_ml4ps}

\usepackage[utf8]{inputenc} % allow utf-8 input
\usepackage[T1]{fontenc}    % use 8-bit T1 fonts
\usepackage{hyperref}       % hyperlinks
\usepackage{url}            % simple URL typesetting
\usepackage{booktabs}       % professional-quality tables
\usepackage{amsfonts}       % blackboard math symbols
\usepackage{nicefrac}       % compact symbols for 1/2, etc.
\usepackage{microtype}      % microtypography
\usepackage{float}
\usepackage{graphicx}
\usepackage{subfigure}

\renewcommand{\d}{\mathrm{d}}
\newcommand{\vS}{{\mathbf{S}}}

\usepackage{lipsum}

\title{Learning the Evolution of the Universe in N-body Simulations}

\author{
  Chang Chen\\
  New York University\\
  \texttt{cc4858@nyu.edu} \\
   \And
  Yin Li\\
  Flatiron Institute\\
  \texttt{yinli@flatironinstitute.org} \\
  \And
  Francisco Villaescusa-Navarro\\
  Princeton University\\
  \texttt{villaescusa.francisco@gmail.com} \\
  \And
  Shirley Ho\\
  Flatiron Institute\\
  \texttt{shirleysurelyho@gmail.com} \\
  \And
  Anthony Pullen\\
  New York University\\
  \texttt{ap177@nyu.edu} \\
}

\begin{document}

\maketitle

\begin{abstract}

Understanding the physics of large cosmological surveys down to small (non-linear) scales will significantly improve our knowledge of the Universe.
Large N-body simulations have been built to obtain predictions in the non-linear regime.
However, N-body simulations are computationally expensive and generate large amount of data, putting burdens on storage.
These data are snapshots of the simulated Universe at different times, and fine sampling is necessary to accurately save its whole history.
We employ a deep neural network model to predict the non-linear N-body simulation at an intermediate time step given two widely separated snapshots.
Our results outperform the cubic Hermite interpolation benchmark method in interpolating N-body simulations. 
This work can greatly reduce the storage requirement and allow us to reconstruct the cosmic history from far fewer snapshots of the universe.

\end{abstract}

\section{Introduction}

Cosmological observables, such as the spatial distribution of galaxies, provide a wealth of information about the fundamental laws and contents of our Universe.
The scientific community has spent billions of dollars in building cosmological surveys that will map different properties of our Universe.
To extract maximum cosmological information from these surveys, precise theoretical predictions of the Universe across cosmic time are needed; however, these are typically hard to come by as they require the full past history of all the photons that have reached us as we observe them. This requires numerical simulations that capture the full temporal evolution of the Universe.

One way to accomplish this objective is through numerical simulations: snapshots of a simulated Universe
can be saved at a large number of frequently-sampled times, allowing us to reconstruct the evolution of the considered observable.
This however will require us to store a large number of the simulation snapshots, which can be very expensive. 
In this work we use a different method to tackle this problem; we propose to use neural networks to interpolate between the output of numerical simulations at two different times. 
This approach will reduce the storage needed for these simulations, and will free up resources that can be used to improve other aspects of the simulations.
% such as volume and/or resolution.
In this work we focus on investigating how well can neural networks learn to interpolate between snapshots from gravity-only (N-body) simulations of the Universe at different cosmological times (redshifts) in 3D.

\section{Method}
\label{method}

In this section we describe the data we use, our model architecture, and the benchmark model.

\paragraph{Data.} Our dataset consists of cosmological $N$-body simulations \cite{1981csup.book.....H} from the \textsc{Quijote} simulation suite \cite{2020ApJS..250....2V}.
Each simulation follows the evolution of $512^3$ particles throughout cosmic history in a box of 1 $(\mathrm{Gpc}/h)^3$ volume (about 5 billion light years in size).
The particles started from locations slightly perturbed from a uniform grid, and later form web-like structures with large density contrast due to gravity (see Fig.~\ref{fig:2d}).
Some particles collapse in dense structures and become very entangled, and full N-body simulations with relatively high resolution is needed to resolve these dense regions. 
Each Quijote simulation saved multiple snapshots at intermediate time steps (redshifts, $z$) before it ends at ``today'' ($z=0$).
We use a total of 101 simulations: 80 for training, 20 for validation, and 1 for testing.

We first associate each particle with two sets of displacements $\mathbf{S}$ and velocity vectors $\mathbf{v}$, each set evaluated at different redshifts; this is our input.
Our target is the displacement vector of each particle at an intermediate redshift.
The displacement vector is defined as $\mathbf{S}=\mathbf{x}-\mathbf{q}$, where $\mathbf{x}$ and $\mathbf{q}$ are the position of a particle at redshift $z$, and its initial position, respectively.
We take advantage of the fact that in the simulations, the initial positions of the particles are laid down in a regular grid, making convolutional neural net an obvious choice for our work here.

To preserve the physical translational equivariance, cosmological simulations are performed with periodic boundary conditions.
For the same reason, convolutional NN needs to adopt the same padding to preserve this symmetry.
However, because of the large number of particles ($512^3$), we cannot feed a whole simulation into the GPU memory during training.
We thus crop the 3D fields of displacements and velocities in regions with $128^3$ voxels.
Using padding from the larger box, we enlarge a bit the input fields ($168^3$ voxels) to make sure the voxels near the boundaries of the target see a significant fraction of their environments.
We made use of data augmentation (rotations and flipping) to force the network to learn the underlying symmetries.

\paragraph{Neural Network Model.}

Since we have organized our data into 3D fields, we can apply a variety of CNNs.
We adopt a U-Net / V-Net type network \cite{unet, vnet}. We use map2map \cite{map2map} as the neural network emulator in our paper.
An important difference from the original U-Net / V-Net architecture is that we add a global bypass connection from the input directly to the output.
The bypass convolves the input by $1^3$ kernel and then directly add the result to the output.
This connection learns a linear interpolation of the input, forcing the rest of the network to learn corrections to it.
Our network uses three stages of resolutions linked in a ``V'' shape taking two downsampling and two upsampling layers.
On each stage residual blocks \cite{resnet} of two $3\times3\times3$ convolutions are inserted to connect input, resampling, and output layers.
We add batch normalization after each convolution except the first one and last two, and use leaky ReLU activation function with negative slope $0.01$.
The channel size is 64 throughout the model except for the input (12), output (3), and after the U-Net concatenation layers (128).

We use the Adam optimizer \cite{adam}, with learning rate 0.0001, $\beta_1, \beta_2 = 0.9, 0.999$, and reduce the learning rate by half when there is no improvement after three training epochs.
We train the neural network to minimize the mean square error (MSE) loss as $\mathcal{L}=\frac{1}{N}\sum_{i=1}^N(\mathbf{S}_i-\mathbf{S}_i^\mathrm{truth})^2$, where $N$ is the total number of particles, $\mathbf{S}_i$ and $\mathbf{S}_i^\mathrm{truth}$, respectively, are the predicted and target displacement vector of the $i$-th particle.

\paragraph{Cubic Hermite Interpolation.}
We compare the results of the neural network against a cubic Hermite interpolator (our benchmark). Given the displacement $\mathbf{S}$ and velocity $\mathbf{v}$ vectors of a given particle at two different redshifts, we can use this interpolator to predict the value of the displacement field at an intermediate redshift.
The coefficients of the polynomial are determined by requiring the polynomial matches $\mathbf{S}$ and its derivative $\d\mathbf{S}/\d z$ at the both input redshifts.
The derivatives are evaluated taking into account that  $\d\vS/\d z=-\mathbf{v}/H(z)$, where $H(z)$ is the Hubble expansion rate of the Universe at redshift $z$. We use the \texttt{scipy} \cite{2020SciPy-NMeth} implementation of cubic Hermite spline interpolation for this procedure.

\section{Results}
\label{results}

We have used the neural network and the cubic Hermite interpolator to predict the $z=1$ displacement field from two snapshots at $z=2$ and $z=0$. In this section we study the performance of the network and its comparison with the benchmark model. We begin with a qualitative visual comparison and then we perform a more detailed quantitative comparison using different summary statistics. We note that we obtain similar results when interpolating to $z=2$ from redshifts $z=3$ and $z=1$, as well as to $z=1$ from $z=2$ and $z=0.5$.

\begin{figure}
\includegraphics[width=0.33\linewidth, height=3.5cm]{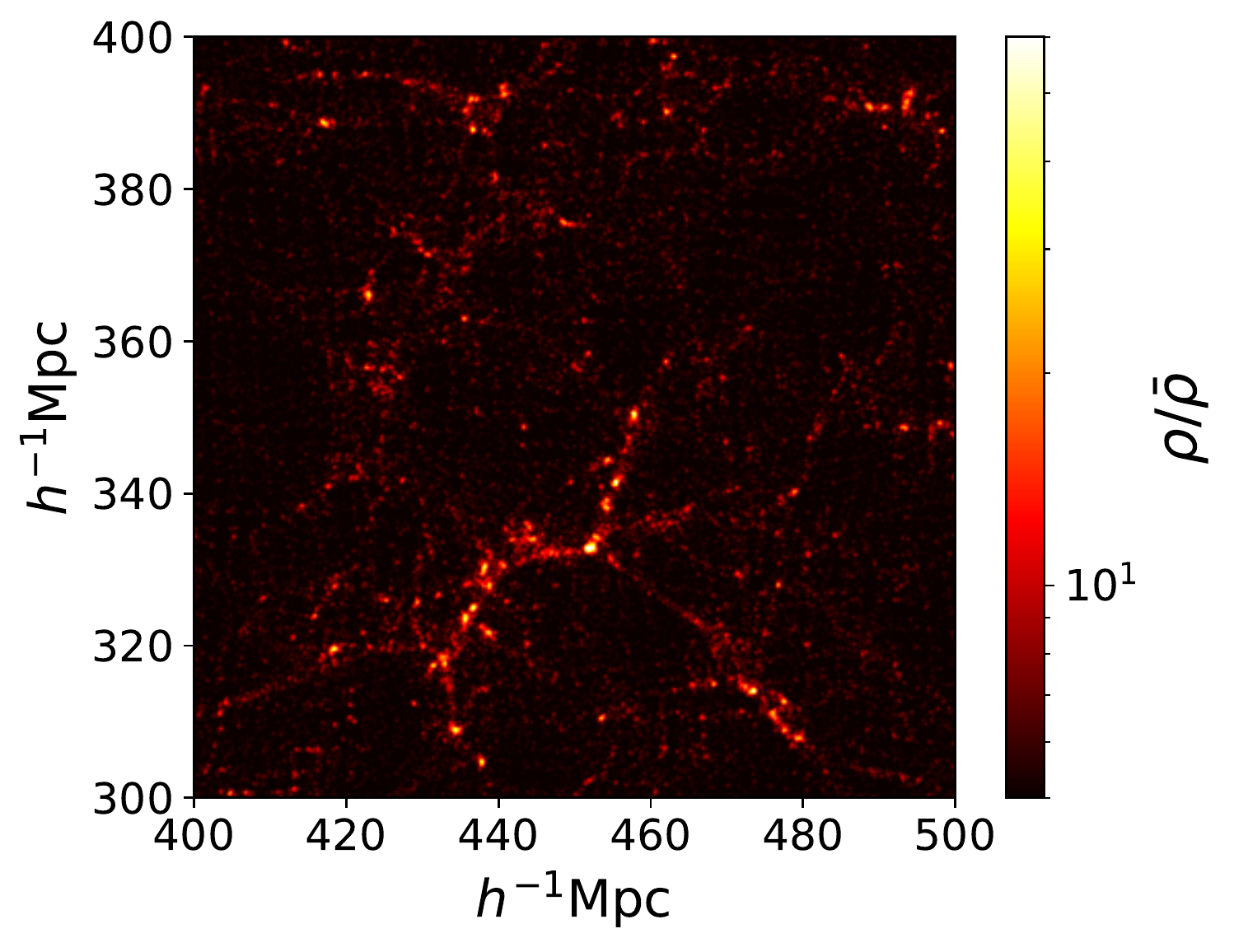}\hfill
\includegraphics[width=0.33\linewidth,height=3.5cm]{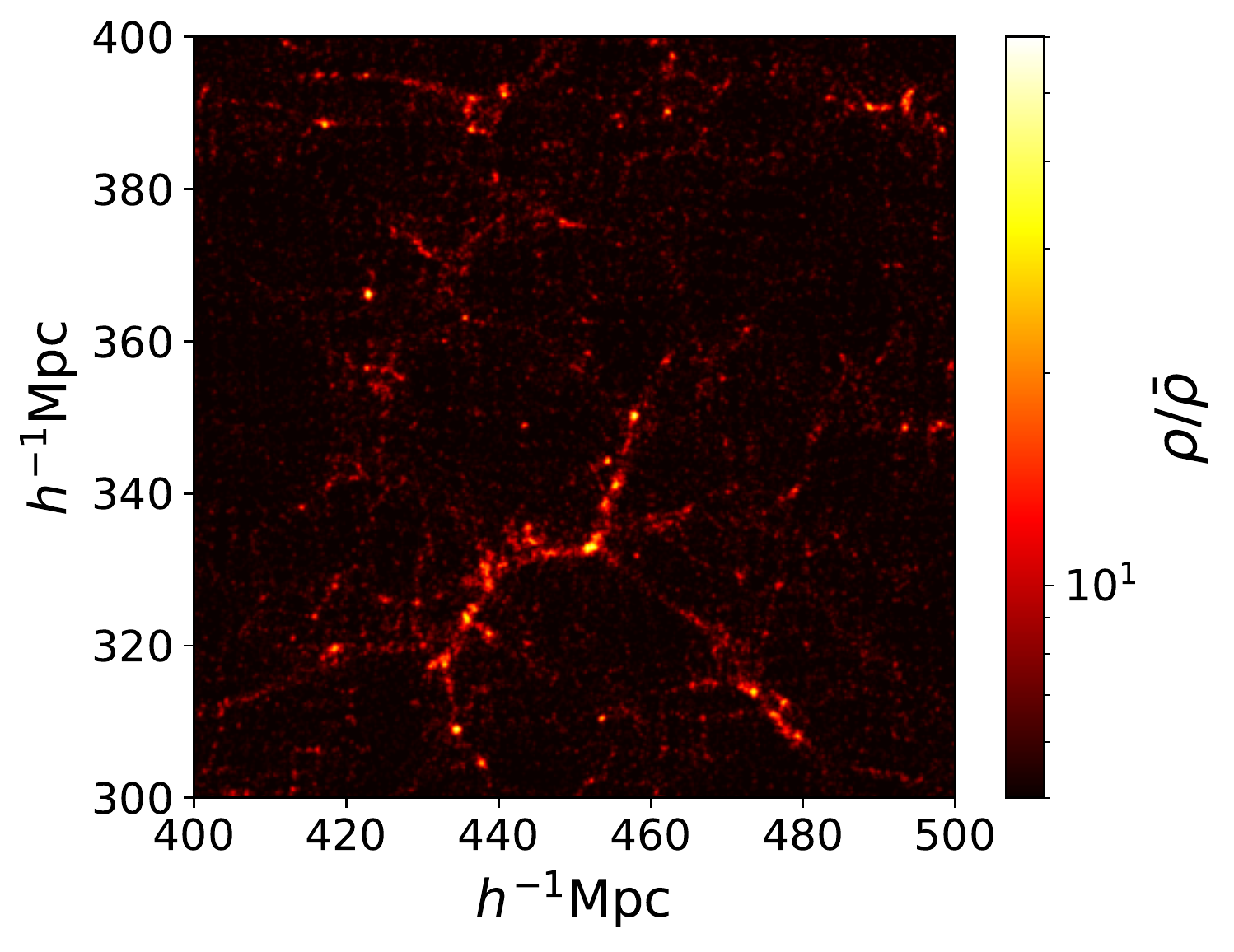}\hfill
\includegraphics[width=0.33\linewidth,height=3.5cm]{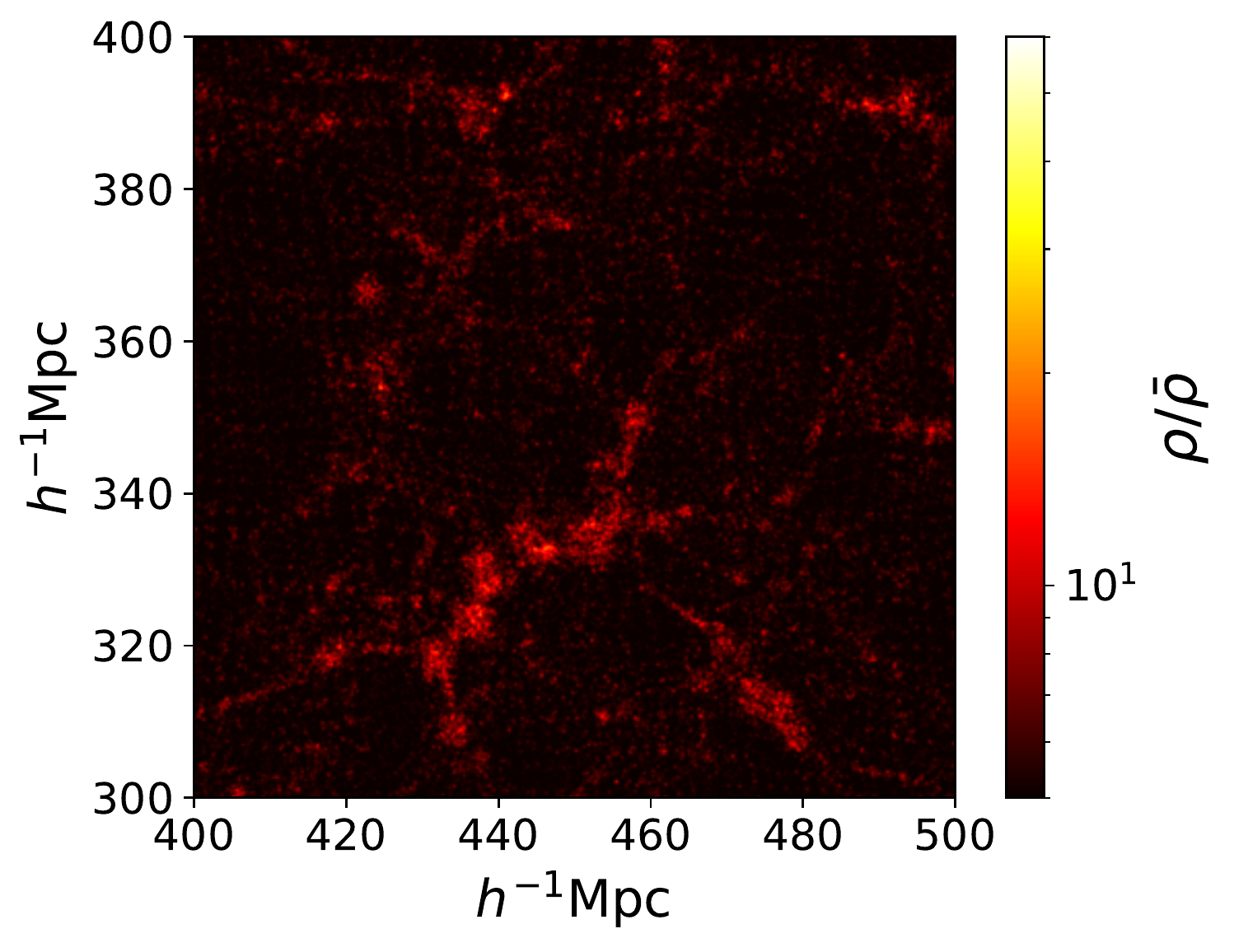}
 \caption{Density projections over a region of $100\times100\times30~(h^{-1}{\rm Mpc})^3$ at $z=1$ for the Quijote simulation (left), V-Net (middle), and cubic Hermite interpolation (right).
 By visual inspection we can see that the NN model performs much better than interpolation.}
\label{fig:2d}
\end{figure}

\paragraph{2D density field.} Fig.~\ref{fig:2d} shows a 2D density projection of one Quijote simulation, and the corresponding region of the NN and interpolation predictions at $z=1$. While the NN model works well on all scales, the cubic Hermite interpolation produces fuzzier structures, specifically the dark matter halos become less concentrated and hence fail to be detected as shown below.

\begin{figure}
\includegraphics[width=0.5\linewidth, height=7cm]{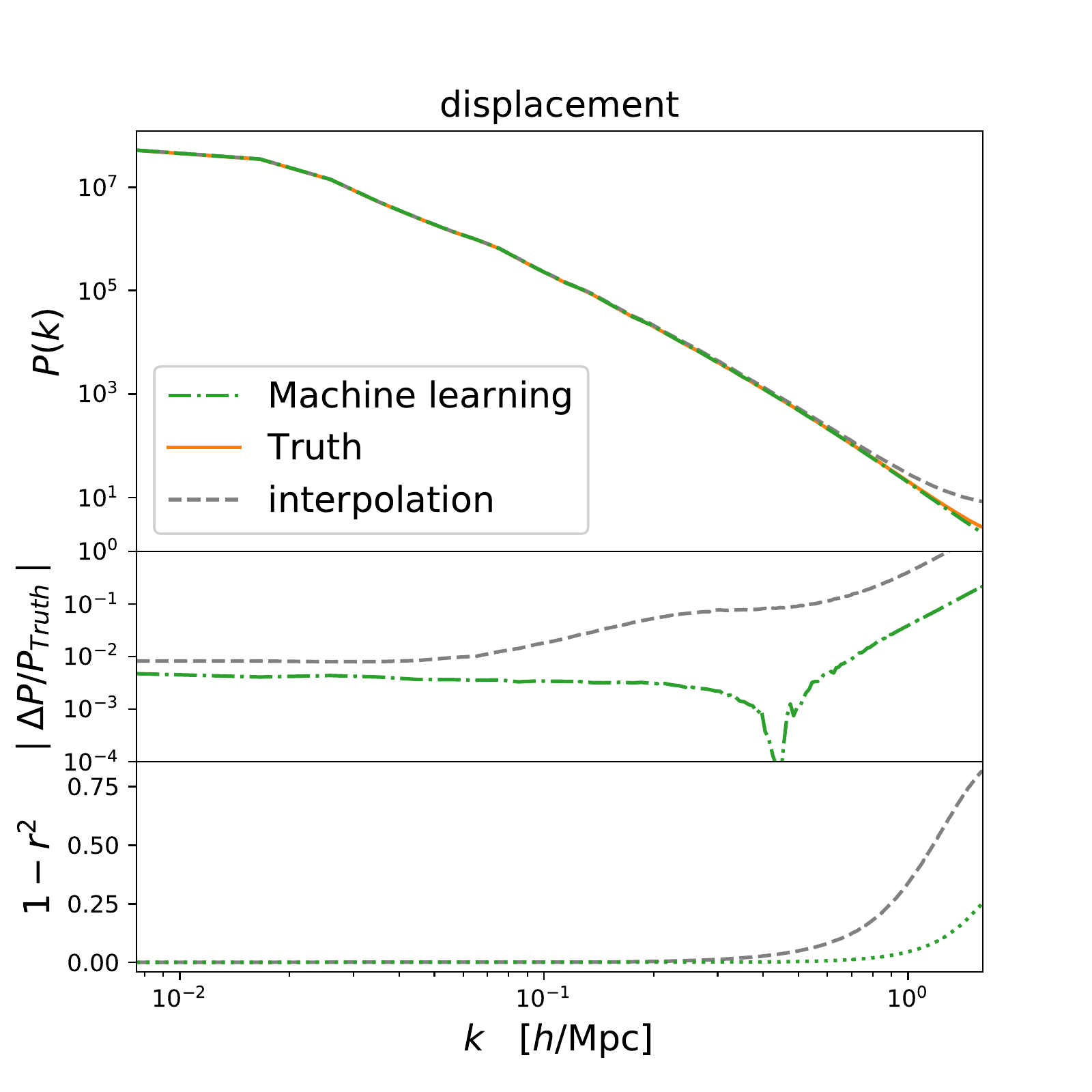}\hfill
\includegraphics[width=0.5\linewidth, height=7cm]{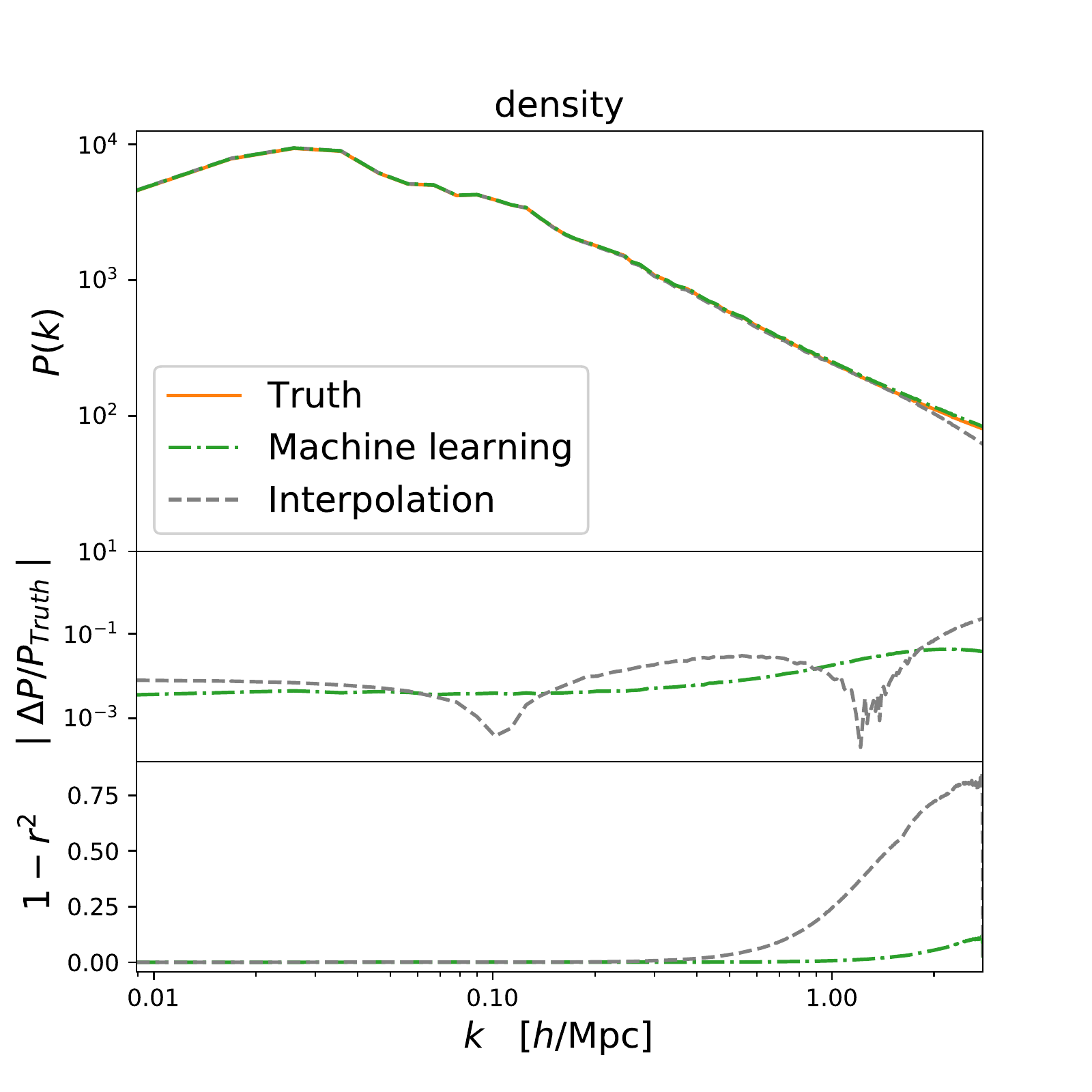}
    \caption{Displacement (left) and density (right) power spectrum (top), relative error (middle) and coefficient of non-determination (bottom) for interpolation from redshift 0 and 2 to 1
}
    \label{fig:power}
\end{figure}

\paragraph{Power spectrum.} One of the most important statistics used to characterize cosmological fields is the power spectrum, that is computed as follows. First, the Fourier transform of the field is calculated. Next, modes are assigned to wavenumber, $k$, bins and the average of their square amplitude is assigned to that $k-$bin. The top panels of Fig. \ref{fig:power} show the power spectra of the displacement and the density fields for a Quijote simulation, the output of the network, and the results of the interpolator. We find that the agreement of the three power spectra are very good. In order to better quantify the agreement, we show in the middle panels $\mid1-P(k)/P_{\rm Truth}(k)\mid$; while the neural network works always performs better than the interpolator for the displacement field, for the density field results are similar, with the exception of very small scales, where the interpolator outperforms the network.

Comparing power spectra inform us on the agreement between mode amplitudes, but neglect the information on phases. In order to quantify the agreement between mode phases for the different methods, we use the Pearson correlation coefficient, defined as $r(k)=\d frac{P_X(k)}{\sqrt{P(k)P_{Truth}(k)} }$, where $P_X(k)$ is the cross-power spectrum between the predicted $P(k)$ the power spectrum from the simulation $P_{Truth}(k)$. The bottom panels of Fig. \ref{fig:power} show the results. In this case, we find that the network performs better than the interpolator at all scales; for the density field, it reaches percent accuracy down to scales as small as $k\sim1~h/{\rm Mpc}$. We note that in general, the accuracy of the network shrinks on small scales (large $k$). This is expected, as this is the the most non-linear regime.

\begin{figure}
\includegraphics[width=0.5\linewidth]{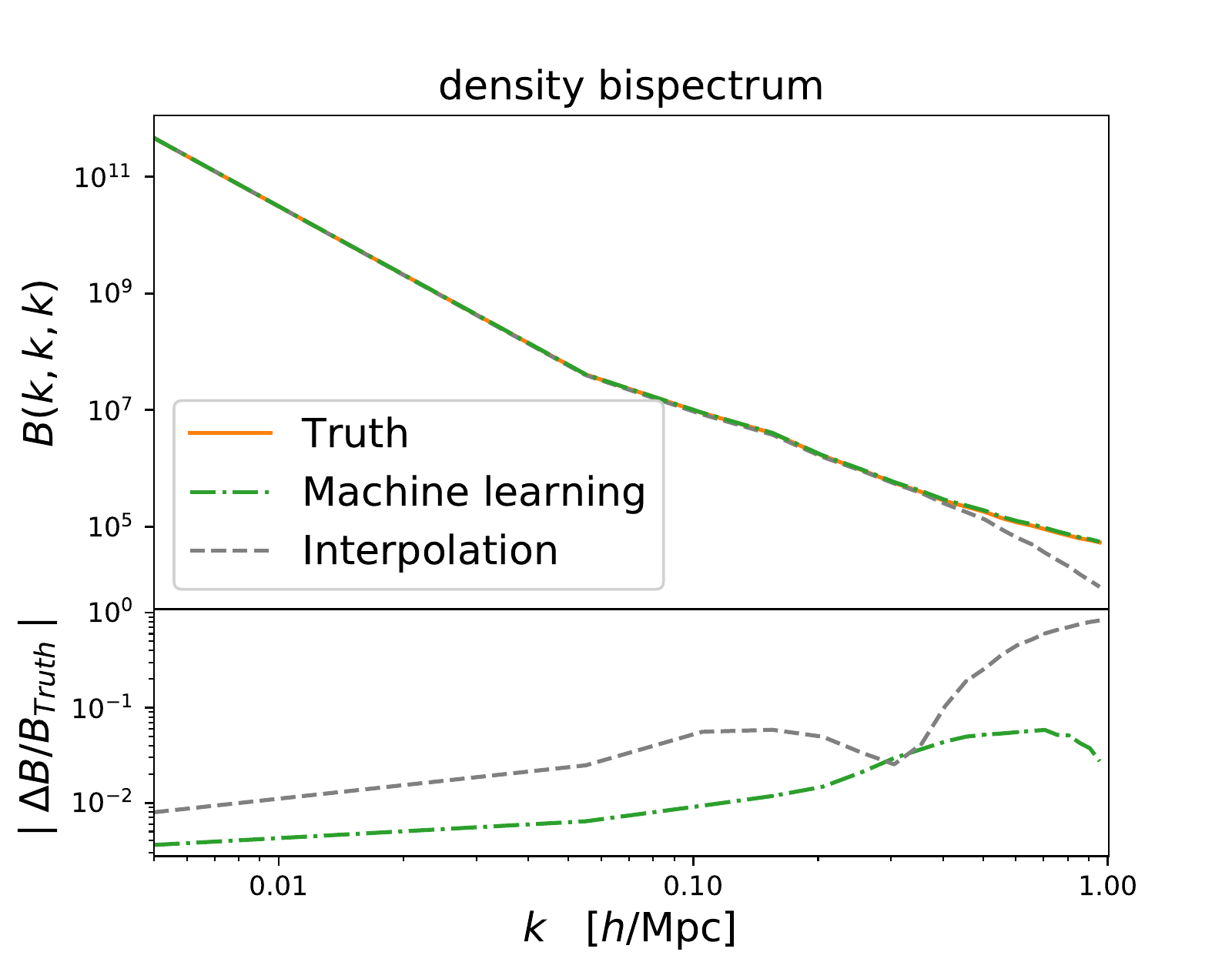}\hfill
\includegraphics[width=0.5\linewidth]{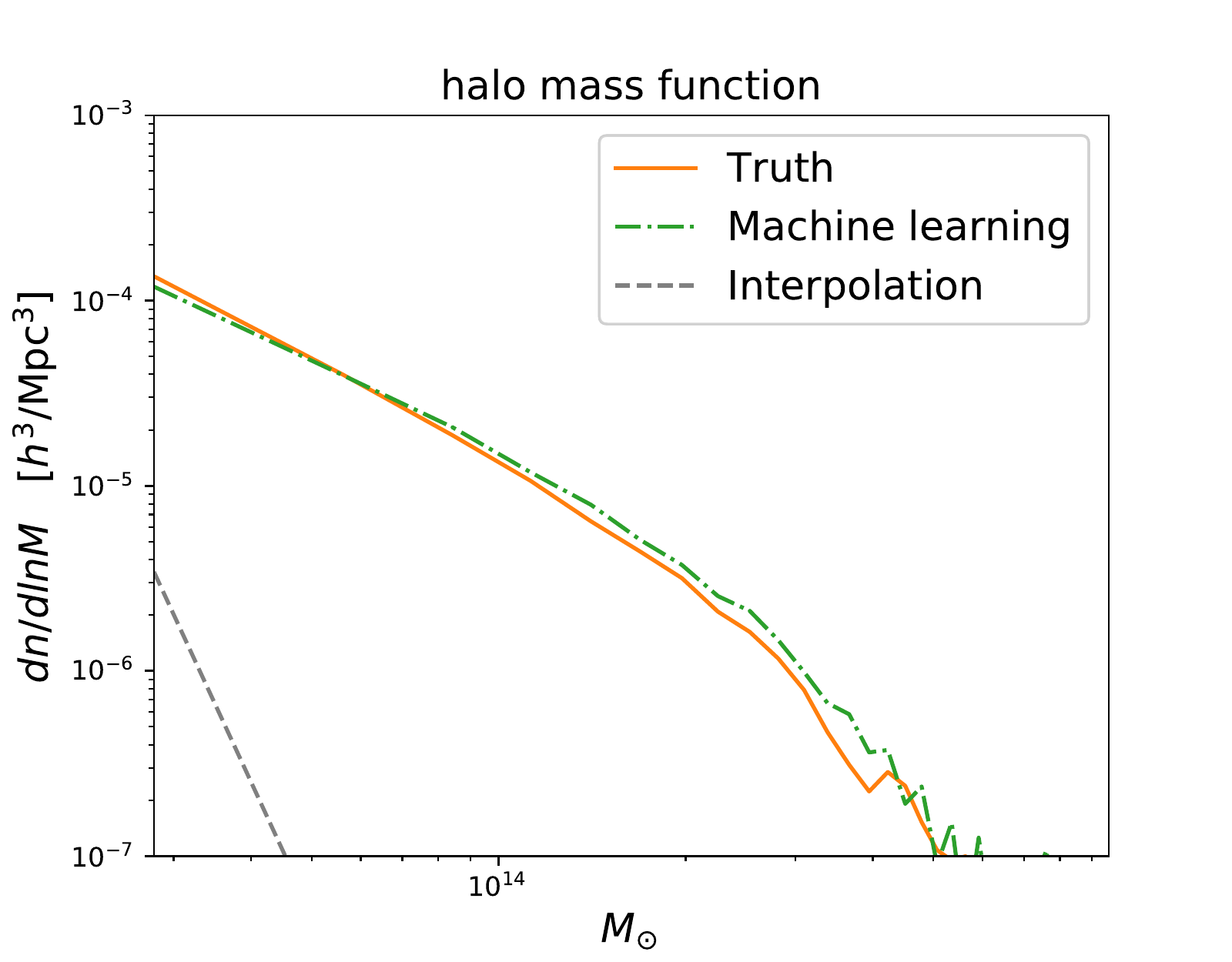}
\caption{Density field equilateral bispectra (left) and halo mass function (right) for interpolation from redshift 0 and 2 to 1.
}
\label{fig:bispectra}
\end{figure}

\paragraph{Bispectrum.} We now consider a higher-order statistic: the bispectrum.  The bispectrum is a statistics that captures information on the field that is different to the one from the power spectrum \cite{Hahn_2019}. We made use of this property to investigate how well the network performs for this statistic. This quantity is calculated by multiplying the amplitude of three modes that form a closed triangle ($\mathbf{k_1}+\mathbf{k_2}+\mathbf{k_3}=0$). As with the power spectrum, $k-$bins are taken and the average of the above product for all modes in that bin is calculated. We have computed the bispectrum of equilateral triangles ($k_1=k_2=k_3=k$) as a function of $k$, and we show the results, normalized by the measurements from the simulation, on the left panel of Fig. \ref{fig:bispectra}. We find that the network achieves a subpercent accuracy down to $k\simeq0.4~h/{\rm Mpc}$, and outperforms the interpolator on all scales.

\paragraph{Dark Matter Halos.} Dark matter halos are gravitationally bound structures (with masses ranging from billions to quadrillions that of the Sun) where galaxies form and live. We have identified these structures in our simulations and in the output of the network and the interpolator using the the FOF (friends-of-friends) halo finder. We have taken a value of $b=0.2$ for the linking length, and only halos with more than twenty particles are considered. We then compute the halo mass function, defined as the abundance of dark matter halos per unit of volume and unit of mass. We use this statistic as it characterizes the 3D density field in a different manner than the power spectrum and bispectrum. We show the results on the right panel of Fig. \ref{fig:bispectra}. We find that the network is able to reproduce the halo mass function of the simulation very accurately. On the other hand, the outcome of the interpolator is completely off. The reason is that, as we saw in Fig. \ref{fig:2d}, the field produced by the intepolator is fuzzier on small scales than the one of the network and the simulation.

\section{Conclusions and future work}

We have shown that neural networks can learn to interpolate between the output of complex numerical N-body simulations that model the dynamical evolution of millions of particles under the force of gravity.
Our V-Net model achieves high accuracy on the four statistics we have investigated in this paper.
Furthermore, the network outperforms
the benchmark model, a cubic Hermite interpolator, on almost all cases.

We plan to extend our work to be able to interpolate snapshots in a continuous manner, i.e. having the value of the interpolated redshift as an input to the network. We will also enhance our network to interpolate not only in redshift, but also as a function of the cosmological parameters. Our work will not only alleviate the computational needs of N-body simulations, but will allow to create more accurate lightcones, needed to analyze the data from cosmological missions.

\section*{Broader Impact}

Our work represents a step forward in the direction of building interpolators that work in field space, rather than in the canonical variables used in traditional interpolators. Our formalism is completely general, and therefore can be easily applied to interpolating between 2D or 3D at different times, or equivalent variables, in other fields. No ethical aspects are relevant for this work.

\begin{ack}

YL thanks the Simons Foundation for support of fellowship. We acknowledge that the work reported on in this paper was partly performed using the Princeton Research Computing resources at Princeton University which is consortium of groups including the Princeton Institute for Computational Science and Engineering and the Princeton University Office of Information Technology's Research Computing department.
We acknowledge that the work was supported in part through the NYU IT High Performance Computing resources, services, and staff expertise.
\end{ack}

\bibliographystyle{unsrt}
\bibliography{references}

\end{document}